\documentclass[english,showpacs,floatfix,amsmath,amssymb,11pt]{revtex4}
\usepackage[dvips]{graphicx}
\usepackage{longtable}
\usepackage{dcolumn}
\usepackage{latexsym}
\usepackage{babel}
\usepackage[latin1]{inputenc}
\usepackage{color}
\usepackage[colorlinks]{hyperref}
\newcommand\tabcaption{\def\@captype{table}\caption}

\def\pa{\partial}

\def\om{\omega}
\def\Om{\Omega}

\def\ga{\gamma}

\def\dl{\delta}

\def\sg{\sigma}

\def\O{O(t,x,y,z)}
\def\S{S(T,X,Y,Z)}
\def\nn{\nonumber}

\def\ln{\mbox {ln}}
\def\wt{\widetilde}
\def\l{\left}
\def\r{\right}

\begin{document}
\title{\Large\bf Some Paradoxes in Special Relativity}
\author{Ying-Qiu Gu}
\email{yqgu@fudan.edu.cn} \affiliation{School of Mathematical
Science, Fudan University, Shanghai 200433, China} 
\date{26th September 2009}

\begin{abstract}
The special theory of relativity is the foundation of modern
physics, but its unusual postulate of invariant vacuum speed of
light results in a number of plausible paradoxes. This situation
leads to radical criticisms and suspicions against the theory of
relativity. In this paper, from the perspective that the
relativity is nothing but a geometry, we give a uniform resolution
to some famous and typical paradoxes such as the ladder paradox,
the Ehrenfest's rotational disc paradox. The discussion shows that
all the paradoxes are caused by misinterpretation of concepts. We
misused the global simultaneity and the principle of relativity.
As a geometry of Minkowski space-time, special relativity can
never result in a logical contradiction.\vskip3mm \noindent
\end{abstract} \maketitle

\section{Introduction}
\setcounter{equation}{0}

In order to reconcile the confliction between Newtonian Mechanics
and Electromagnetism, in 1905 Einstein introduced the special
relativity by two postulates, that is, the vacuum light speed is
independent of observer and the Galilean principle of relativity.
The first postulate is so unusual that a number of plausible
paradoxes are long standing without a resolution of consensus. The
suspicions that the relativity may be contradictory always exist
among physicists. In the non-mainstream literatures, we can see a
lot of radical criticisms against the theory of relativity.

Historically, we can find the profound theoretical and
experimental origin of relativity. In the late 19th century, Henri
Poincar\'e suggested that the principle of relativity holds for
all laws of nature. Joseph Larmor and Hendrik Lorentz discovered
that Maxwell's equations were invariant only under a certain
change of time and length units. This left quite a bit of
confusion among physicists, many of whom thought that a
luminiferous aether is incompatible with the relativity principle.

In their 1905 papers on electrodynamics, Henri Poincar\'e and
Albert Einstein explained that with the Lorentz transformations
the relativity principle holds perfectly. Einstein elevated the
principle of relativity to an axiom of the theory and derived the
Lorentz transformations from this principle combined with the
principle of constant vacuum speed of light. These two principles
were reconciled with each other by a re-examination of the
fundamental meanings of space and time intervals\cite{relat}.

However, the physical description of the constant light speed is
ambiguous and looks contradictory in the common sense. What is the
light speed? why should it be related to the vacuum? and how can a
speed be independent of moving frames are the usual complains and
arguments. Such ambiguous statement is not easy to be clarified in
physics, which involves a number of basic concepts and theories.
It even leads to misinterpretation and puzzles among sophistic
experts. As a matter of fact, the speed of light $c$ has
complicated physical background and meanings involving the mystic
properties of electromagnetic field, and is hidden in almost all
equations but with different facets\cite{Ellis}. As the most
fundamental concepts or principles in physics, the simpler the
meanings and expressions, the better the effectiveness and
universality.

In essence, the relativity is nothing but a 4-dimensional
geometry. The confusion and paradoxes are caused by conceptual
misinterpretation. In this paper, we give a unified resolution to
some famous paradoxes according to the underlying geometry of the
special relativity. The following typical paradoxes in relativity
and some versions of their resolutions can be found in the
Wikipedia free encyclopedia and some textbooks.

\begin{enumerate}

\item {\bf Twins paradox}\cite{twins}:  The twin paradox is a
thought experiment in special relativity, in which a twin who
makes a journey into space in a high-speed rocket will return home
to find he has aged less than his identical twin who stayed on
Earth. This result appears puzzling on this basis: the laws of
physics should exhibit symmetry. Either twin sees the other twin
as travelling; so each should see the other aging more slowly. How
can an absolute effect (one twin really does age less) result from
a relative motion?

\item {\bf Bell's spaceship paradox}\cite{bell}: It is a thought
experiment in special relativity involving accelerated spaceships
and strings. Two spaceships, which are initially at rest in some
common inertial reference frame are connected by a taut string. At
time zero in the common inertial frame, both spaceships start to
accelerate, with a constant proper acceleration as measured by an
on-board accelerometer. Question: does the string break? i.e.,
does the distance between the two spaceships increase?

\item {\bf The ladder paradox}\cite{ladder} The ladder paradox(or
barn-pole paradox) is a thought experiment in special relativity.
If a ladder travels horizontally it will undergo a length
contraction and will therefore fit into a garage that is shorter
than the ladder's length at rest. On the other hand, from the
point of view of an observer moving with the ladder, it is the
garage that is moving and the garage will be contracted. The
garage will therefore need to be larger than the length at rest of
the ladder in order to contain it. Suppose the ladder is
travelling at a fast enough speed that, from the frame of
reference of the garage, its length is contracted to less than the
length of the garage. Then from the frame of reference of the
garage, there is a moment in time when the ladder can fit
completely inside the garage, and during that moment one can close
and open both front and back doors of the garage, without
affecting the ladder. However, from the frame of reference of the
ladder, the garage is much shorter than the length of the
ladder(it is already shorter at rest, plus it is contracted
because it is moving with respect to the ladder). Therefore there
is no moment in time when the ladder is completely inside the
garage; and there is no moment when one can close both doors
without it hitting the ladder. This is an apparent paradox.

\item {\bf Ehrenfest paradox}\cite{disc}: The Ehrenfest paradox
concerns the rotation of a ``rigid'' disc in the theory of
relativity. In its original formulation as presented by Paul
Ehrenfest 1909 in the Physikalische Zeitschrift, it discusses an
ideally rigid cylinder that is made to rotate about its axis of
symmetry. The radius $R$ as seen in the laboratory frame is always
perpendicular to its motion and should therefore be equal to its
value $R_0$ when stationary. However, the circumference $2\pi R$
should appear Lorentz-contracted to a smaller value than at rest,
by the usual factor $\ga=\sqrt{1-v^2}$. This leads to the
contradiction that $R=R_0$ and $R<R_0$.

\end{enumerate}

\section{Geometry of the Minkowski Space-time}
\setcounter{equation}{0}
\subsection{Basic Concepts and Settings}

Like the Euclidean geometry, if we directly express the special
relativity as geometry of the Minkowski space-time at the
beginning, and then explain the corresponding rules between the
geometrical concepts and the reality, the problem becomes clearer
and simpler. Mathematically special relativity is a simple
theory\cite{Vankov}. In some sense, the postulate of constant
light speed is a restriction on the metric of the space-time, and
the principle of relativity explains the rules of transformation
between coordinate systems. The following contents are some well
known knowledge, but we organize them consistently in logic.

The events or points in the space-time can be described by 4
independent coordinates. This is a basic fact and can be only
accepted as a fundamental assumption\cite{gu1,matt1}. A more
profound philosophical reason for the 1+3 dimensions is that, the
world is related to the elegant mathematical structure of the
quaternion or Clifford algebra as shown below. Assume $S(T, X ,Y,
Z)$ is a static coordinate system, where the physical meanings of
the coordinates $(T, X ,Y, Z)$ is that, $T$ is the period counting
of an idealized oscillator without any interaction with other
matter, and $(X ,Y, Z)$ is measured by an idealized rigid unit
box. We imply that such measurement can be performed everywhere,
and the values assign a unique coordinate to each event in the
space-time. However, this operation is not enough to describe the
property of the space-time, because the space-time is measurable,
in which the distance between two events is independent of
measurement. Similar to the Euclidean geometry, the space-time
manifold has a consistent distance $ds$ between events. Of course,
it is just an assumption in logic and should be tested by
experiments. According to the electrodynamics, which is mature
theory proved by numerous experiments, we find that in the Nature
the consistent distance should be quadratic form $ds^2=g_{\mu\nu}
dx^\mu dx^\nu$. In special relativity, which implies the
space-time is uniform and flat, then we have
\begin{eqnarray}
ds^2= dT^2 -dX^2-dY^2-dZ^2,\qquad \forall(T,X,Y,Z), \label{2.1}
\end{eqnarray}
where and hereafter we set $c=1$ as unit if it is not confused.
Sometimes we use $\vec X= (X^1,X^2,X^3)$ and so on to denote the
3-dimensional vector.

(\ref{2.1}) is nothing but a kind of Pythagorean theorem, which
defines the Minkowski space-time. (\ref{2.1}) is the starting
point of the following discussions. Although (\ref{2.1}) is
compatible with the postulate of constant light speed, it gets rid
of the definition of light speed $c$, and becomes clear and simple
in logic. The explanation of $c$ is related with the complicated
properties of electromagnetic field which should be explained
according to electrodynamics.

The distance formula (\ref{2.1}) and the dynamical equations in
the space-time all have the elegant structure of quaternion or the
Clifford algebra $C\ell_{1,3}$\cite{clf1,clf2,clf3,clf4,gu3}.
Assume $\ga^\mu$ is a set of basis satisfying the fundamental
relation of the Clifford algebra
\begin{eqnarray}
\ga^\mu\ga^\nu+\ga^\mu\ga^\nu=2g^{\mu\nu},\qquad \mu,\nu
\in\{0,1,2,3\},  \label{*2.1}
\end{eqnarray}
which can be realized by the Dirac matrices, then the vector in
the Minkowski space-time can be expressed by $d\textbf{X}=dX^\mu
\ga_\mu\in C\ell_{1,3}(\mathbb{R})$, and (\ref{2.1}) simply
becomes the geometric product $ds^2=d\textbf{X}^2$. The geometric
product of any vectors $\textbf{a}$ and $\textbf{b}$ is defined by
$\textbf{a}\textbf{b}=\textbf{a}\cdot\textbf{b}+\textbf{a}\wedge\textbf{b}.$
And the dynamical equation of all realistic fields
$\Psi=(\psi_1,\psi_2,\cdots,\psi_n)$ becomes $\pa \Psi =f(\Psi)$,
where $\pa=\ga^\mu\pa_\mu$ is quaternion differential operator,
and $f(\Psi)$ consists of some tensorial products of $\Psi$ such
that the dynamical equation is covariant\cite{matt1,gu3}.

From (\ref{2.1}), we can derive a series of equivalent coordinate
system according to the principle of relativity, in which the
distance of the space-time also takes the form (\ref{2.1}).
Obviously, such coordinate transformation must be linear, which
forms the Poincar\'e group. This means the new coordinate systems
should be constructed by the similar operational procedure, which
give the definition of inertial coordinate systems. The
transformation is generally given by
\begin{eqnarray}
d\textbf{x}\equiv dx^\mu\ga_\mu =\ga^0\Om d\textbf{X} \Om\ga^0,
\label{*2.2}
\end{eqnarray}
where $\Om=\om^\mu\ga_\mu$ are any given vector satisfying
$\Om^2=1$. By (\ref{*2.2}), we can easily check the following
results: (I). $d\textbf{x}^2=d\textbf{X}^2$; (II). $\Om$
corresponds to the boost transformation if $\vec \om\in
\mathbb{R}^3$, and to the rotating transformation if $\vec \om\in
\mathbb{C}^3$.

The rotating transformation has clear geometric meaning, and is
well understood. The paradoxes only involves the boost
transformation, so we take
\begin{eqnarray} \Om=\cosh\frac \xi 2\ga_0 +\sinh\frac\xi 2\ga_1
\end{eqnarray}
as example to show the problem. In this case, expanding
(\ref{*2.2}) we get the usual Lorentz transformation
\begin{eqnarray}
T =t \cosh\xi +x\sinh\xi  ,\quad X = t\sinh\xi + x\cosh\xi ,\quad Y = y,\quad Z = z, \label{2.2}\\
t =  T \cosh\xi  - X\sinh\xi ,\quad x =  X\cosh\xi -
 T \sinh\xi,\quad y = Y,\quad z = Z, \label{2.3}
\end{eqnarray}
where $O(t,x,y,z)$ stands for the new coordinate system. The two
coordinate systems describe the same space-time, and each one
assign a unique coordinate to every event or point of the
space-time. (\ref{2.2}) and (\ref{2.3}) is the 1-1 mapping between
two coordinate systems for each point.

Now we examine the relation between two coordinate system, and
physical meaning of parameter $\xi$. For the origin of
$O(t,x,y,z)$, namely, the point $x=y=z=0$, by (\ref{2.3}) we get
relations
\begin{eqnarray}
&X=VT,\qquad Y=0,\qquad Z=0,& \label{2.4}\\ &t = \sqrt{1-V^2}T,&
\label{2.5}
\end{eqnarray}
where $V=\tanh\xi$, and then we have
\begin{eqnarray}
\cosh\xi =\frac 1 {\sqrt{1-V^2}},\qquad \sinh\xi=\frac V
{\sqrt{1-V^2}}. \label{2.6}\end{eqnarray} The kinematical meaning
of (\ref{2.4}) is obvious, that is, the origin $(0,0,0)$ of $\O$
moves along $X$ at speed $V$ with respect to $\S$. So the new
coordinate system $O(t,x,y,z)$ is a reference frame moves at
relative speed $V$ with respect to $S(T,X,Y,Z)$.

However, the relation (\ref{2.5}) is unusual and counterintuitive,
which means the moving oscillator slows down with respect to the
static one. On the other hand, the static clock also moves with
respect to $O(t,x,y,z)$, and then the static clock should also be
slower than the clock in $O(t,x,y,z)$. This is clock paradox,
which is the origin of the twin paradox. From the derivation of
(\ref{2.5}) we find that two times $T$ and $t$ have different
physical position. $T$ is the global time in $S(T,X,Y,Z)$, but $t$
is just a local time in the viewpoint of the observer at the
static coordinate system $\S$. The hyperplane of simultaneity does
not generally exist for all moving frame\cite{gu}. Now we give
more discussions on this problem.

\begin{figure}[h]
\centering
\includegraphics[width=17cm]{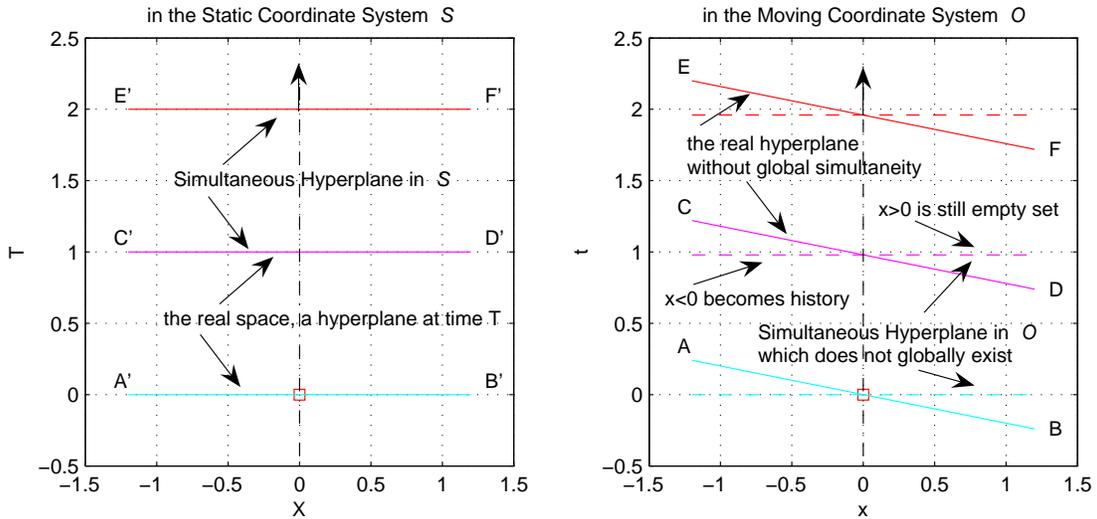}
\caption{The real space of the world is an evolving hyperplane.
Only in one class of special coordinate system we have global
simultaneity.} \label{fig}
\end{figure}
From FIG.(\ref{fig}) we find that, only in one special coordinate
system, the real world can define the concept of global
simultaneity, otherwise it will result in contradiction. We show
this by detailed discussion. Assume that the global simultaneity
holds in coordinate system $S$, that is the world evolves from
$T=0$ to $T=T_0$ as shown in FIG.(\ref{fig}). In this case,
according to the Lorentz transformation (\ref{2.3}), the real
space evolves from the hyperplane $AB$ to $CD$ in $O$, which is a
tilting line without simultaneity. If we toughly define the
simultaneity $t=t_0$ in system $O$, we find that we actually erase
the history in the region $x<0$, and fill up the future in the
region $x>0$. Of course such treatment is absurd, because the
evolution of the world is even not uniquely determined, and we can
neither exactly forecast the future, nor change the history of the
past. In this sense, the hyperplanes $T=T_0$ and $t=t_0$ describe
two different worlds. This evolving characteristic of Lorentz
manifold is essentially different from that of Riemann manifold.
This distinction is similar to the characteristics of the
solutions to the hyperbolic partial differential equation and
elliptic one. The speciality of the temporal coordinate can also
be identified from the form of the Clifford algebra (\ref{*2.2}).

The above descriptions briefly explain the intrinsic property of
the Minkowski space-time and define the inertial reference frames,
which get rid of the complicated explanation of constant light
speed. The constant light speed can be derived from
electrodynamics compatible with the Minkowski space-time. The
equations (\ref{2.1})-(\ref{2.3}) are the fundamental relations in
Minkowski space-time, and all other properties can be derived from
these relations. The following discussions are only based on these
relations and the principle of relativity.

\subsection{Kinematics of a Particle}

The motion of one point (i.e. a particle) is described by
$(T(p),X(p), Y(p),Z(p))$ in coordinate system $S(T,X,Y,Z)$, where
$p$ is the time-like parameter of the world line of the particle.
Usually we set $p=T$ or $p=\tau$(the proper time defined below).
In the following discussion, we assume that in the reference frame
$\S$ we realistically have the global simultaneity $T=T_0$. The
velocity of the particle $v$ with respect to frame $\S$ is defined
by
\begin{eqnarray}
\vec v=\l(\frac {d X}{dT},\frac {d Y}{dT},\frac {d Z}{dT}\r)=\frac
{d\vec R}{dT}, \label{2.7}
\end{eqnarray}
where $(d\vec R,dT)$ means the coordinate elements on the world
line. We define the proper time of the particle as $d\tau\equiv
ds$. Substituting (\ref{2.7}) into (\ref{2.1}), we get
\begin{eqnarray}
d\tau=\sqrt{1-v^2} dT, \label{2.8}
\end{eqnarray}
where $\vec v=\vec v(T)<1$ is an arbitrary continuous function.

In a moving reference frame $O(t,x,y,z)$, the speed with respect
to $\O$ is defined by $\vec u=\frac {d }{dt}\vec r$. According to
the coordinate transformation rule (\ref{2.3}), we get
\begin{eqnarray}
{dt(T)} = ({ \cosh\xi  - v_x \sinh\xi}){dT }= \frac {1-v_x V}
{\sqrt{1-V^2}}{dT }, \label{2.9}
\end{eqnarray}
then the transformation law of speed between two coordinate system
becomes
\begin{eqnarray}
u_x &=& \frac {v_x \cosh\xi -
  \sinh\xi}{ \cosh\xi  - v_x \sinh\xi}= \frac {v_x  -
 V}{ 1  - v_x V} ,\label{1.9}\\
u_y &=& \frac {v_y}  {\cosh\xi  - v_x\sinh\xi}=\frac {\sqrt{1-V^2}
} { 1 - v_x V}v_y , \label{1.10} \\  u_z &=& \frac {v_z} {\cosh\xi
- v_x\sinh\xi}=\frac {\sqrt{1-V^2} } { 1  - v_x V}v_z.
 \label{1.11}
\end{eqnarray}
By (\ref{2.9})-(\ref{1.11}), we can check
\begin{eqnarray}
\sqrt{1-u^2}=\frac{\sqrt{1-v^2}} { \cosh\xi  - v_x
\sinh\xi}=\sqrt{1-v^2}\frac {\sqrt{1-V^2}}{1-v_x
V}=\sqrt{1-v^2}\frac {dT }{dt},
 \label{1.12}
\end{eqnarray}
or simply,
\begin{eqnarray}
\sqrt{1-u^2}dt=\sqrt{1-v^2}dT.
 \label{1.12.1}
\end{eqnarray}
(\ref{1.12.1}) can be also derived from (\ref{2.1}) by the
definition of speed, which implies the proper time between two
definite events (\ref{2.8}) is an invariant quantity independent
of coordinate system and the moving state of the particle. It is
the intrinsic time homogenously recorded by the idealized
oscillator carried by the particle. So the time dilation is
independent of whether the particle is accelerating or moving at
constant speed. More formally, we have

{\bf Theorem 1.} {\em The proper time $\tau$ of a particle moves
at speed $\vec u(t)$ with respect to an inertial coordinate system
$O(t,x,y,z)$ can be generally calculated by
\begin{eqnarray}
\tau=\int_0^t \sqrt{1-u^2(t)}dt.
\end{eqnarray}
$\tau$ is an invariant scalar independent of whether $u(t)$ is a
constant or not, and what inertial coordinate system is referred
to. It is the intrinsic time of the particle.}

If we take the proper time $\tau$ as the parameter of the world
line of the particle, we can define the 4-dimensional speed
$U^\mu$ and acceleration $a^\mu$ of the particle as follows,
\begin{eqnarray}
U^\mu \equiv \dot X^\mu=\frac 1 {\sqrt{1-v^2}}(1,\vec v),\qquad
a^\mu \equiv \dot U^\mu,
\end{eqnarray}
where the over dot stands for $\frac {d}{d\tau}$. Or in the form
of Clifford algebra
\begin{eqnarray}
\textbf{U}= \dot X^\mu\ga_\mu,\qquad \textbf{a}= \dot
U^\mu\ga_\mu.
\end{eqnarray}
By the invariance of $\tau$ and (\ref{2.3}), we get

{\bf Theorem 2.} {\em Under coordinate transformation (\ref{2.2})
and (\ref{2.3}), the transformation law for the 4-dimensional
speed is given by
\begin{eqnarray}
u^0 =  U^0 \cosh\xi  - U^1\sinh\xi ,\quad u^1 =  U^1\cosh\xi -
 U^0 \sinh\xi,\quad u^2 = U^2,\quad u^3 = U^3, \label{2.17}
\end{eqnarray} where $U^\mu$ and $u^\mu$ are respectively 4-dimensional speed
of a particle relatively to the coordinate system $S(T,X,Y,Z)$ and
$O(t,x,y,z)$. Or in the general form of Clifford algebra under
(\ref{*2.2})}
\begin{eqnarray}
\textbf{u}\equiv u^\mu\ga_\mu =\ga^0\Om \textbf{U} \Om\ga^0.
\label{*2.3}
\end{eqnarray}

By the principle of relativity, (\ref{2.17}) or (\ref{*2.3})
actually holds for any 4-dimensional (contravariant) vector such
as $a^\mu$. It should be pointed out that, the coordinate system
and observer are different concepts which are sometimes confused.
A concrete coordinate system is the language for the observer to
express his laws and results of measurement, and it is a
mathematical setting of the observer for the space-time rather
than the observer himself.

\section{the paradoxes for particles}
\setcounter{equation}{0}

\subsection{the Twins Paradox}

The twins paradox involves the computation of the proper time of
two particles in the Minkowski space-time. Starting with Paul
Langevin in 1911, there have been numerous explanations of this
paradox, all based upon there being no contradiction because there
is no symmetry: only one twin has undergone acceleration and
deceleration, thus differentiating the two cases. This is the
standard explanation and consensus\cite{twins}. There are also
some other explanations such as using the radar time to define the
hypersurface of simultaneity for a observer, where the
hypersurface depends on the future world line of the observer, so
it seems quite ambiguous and complicated. We can not confuse the
coordinate system with the observer himself or the object
observed.

In fact, if we measure the proper time of the twins in any moving
inertial coordinate system $\O$, and the result is the same as in
$\S$, then the paradox vanishes.  In mathematics, the theorems 1
and 2 actually imply the results should be the same. For
clearness, in what follows, we make a detailed computation in a
moving coordinate system and compare the results. Such calculation
can clearly show how the subtle concepts of coordinates and
principle of relativity works, and the comparison will dismiss any
ambiguity and suspicion.

Assume in $\S$ the coordinates of the world lines of the static
and the travelling twins are respectively given by
\begin{eqnarray}
\bar X^\mu=(T,0,0,0),\qquad X^\mu=(T,X(T),Y(T),Z(T)).\label{3.0}
\end{eqnarray}
The travelling twin takes off at $T=0$, and return home at
$T=T_0$. According to the definition of the proper time, we have
\begin{eqnarray}
\bar\tau &=& \int_0^{T_0} dT=T_0, \label{3.3.0}\\
\tau &=& \int_0^{T_0} \sqrt{d T^2- d X(T)^2} =
\int_0^{T_0}\sqrt{1-v^2(T)}dT . \label{3.4.0}
\end{eqnarray}
In order to calculate the proper time in moving $\O$, we must
measure the coordinates of the world lines of the twins in $\O$.
So the relation between the coordinates in two coordinate systems
is the key of the paradox. Of course the coordinates are connected
by the Lorentz transformation (\ref{2.2}) and (\ref{2.3}). {\bf No
matter whether we obtain the coordinates of world lines by the
practical measurement or by transformation, the results should be
the same}. This is the essential implication of the principle of
relativity. Substituting (\ref{3.0}) into (\ref{2.3}), we get the
coordinates of the two twins measured by observer in $\O$
respectively by
\begin{eqnarray}
&\bar t(T) =   T \cosh\xi,  \quad \bar x(T) =  -  T
\sinh\xi,\quad \bar y = \bar z = 0, \quad T\in(0,T_0) &\label{3.1}\\
&t(T) =  T \cosh\xi  - X(T)\sinh\xi ,~~ x(T) =  X(T)\cosh\xi - T
\sinh\xi,~~ y = z = 0,~~& \label{3.2}
\end{eqnarray}
where $T$ acts as parameter. Then the proper time of the twins
$\bar\tau$ and $\tau$ measured in $\O$ should respectively be the
following
\begin{eqnarray}
\bar\tau &=& \int \sqrt{d\bar t^2- d\bar x
^2}=\int_0^{T_0}\sqrt{\cosh^2\xi-
\sinh^2\xi} dT =T_0, \label{3.3}\\
\tau &=& \int \sqrt{d t^2- d x ^2} \nn\\
&=& \int_0^{T_0} \sqrt{ (\cosh\xi - v(T) \sinh\xi)^2-(
v(T)\cosh\xi -
\sinh\xi)^2}dT\nn\\
&=& \int_0^{T_0}\sqrt{1-v^2(T)}dT . \label{3.4}
\end{eqnarray}
Comparing (\ref{3.3})-(\ref{3.4}) with
(\ref{3.3.0})-(\ref{3.4.0}), we learn that the proper time is
indeed independent of coordinate system, and we still have
$\bar\tau>\tau$. The clock of the travelling twin slows down. Here
we can not confuse the proper time $\tau$ with the time coordinate
$t$.

In the above integrals, the events that ``The travelling twin
takes off at $T=0$, and return home at $T=T_0$'' is important.
Only under this condition, the region of integration $(0,T_0)$ has
definite meaning independent of coordinate system. This condition
implies the travelling twin moving along a closed spacial curve at
varying speed. This paradox has nothing to do with the curved
space-time. However, if we accept the fact that only in one
coordinate system (strictly speaking, it is a class of coordinate
systems connected by transformation of translation and rotation)
we have global simultaneity, which defines the cosmic time acting
as an absolute standard, then such condition can be cancelled.

The uniqueness of simultaneity means {\bf the World is just one
solution of the physical laws}, so it does not contradict the
principle of relativity\cite{gu}. In what follows, we have some
other examples to deal with the unique global simultaneity and
covariance consistently. The resolution of the
Rietdijk-Putnam-Penrose's Andromeda paradox also needs the
uniqueness of simultaneity in a world\cite{andro}.

\subsection{the Spaceship Paradox}

The Bell's spaceship paradox involves the computation of the
synchronous distance between two particles in different coordinate
system. Assume two spaceships $A$ and $B$ have the same structure,
and they are at rest in the static coordinate system $\S$ when
$T\le 0$. They are linked by an elastic string. Each spaceship
carries a standard clock adjusted to be synchronous before taking
off and have the same operational schedule of engine. Each
spaceship can be treated as a point in $\S$. The following
calculations basically agree with that of Dewan \& Beran and also
Bell\cite{bell,bell1,bell2}.

According to their operation, both spaceships should have the same
proper acceleration (i.e. 4-dimensional form) $a$ along $X$. Then
the equation of motion for any one spaceship is given by
\begin{eqnarray}
\frac d {dT} X&=&v(T),\\
\frac d {d\tau} u^1&=&\frac {1}{\sqrt{1-v^2}}\frac d {d T} \frac
{v}{\sqrt{1-v^2}}=a(\tau), \label{3.6}\end{eqnarray} in which
$u^1$ is the 4-vector speed, $a(\tau)$ is a given function
determined by operational schedule of engine. Theoretically, we
can solve the solution $(v(T),X(T))$ from the equations with
initial values. Usually the solutions can not be expressed by
elementary functions, but they certainly take the following form
\begin{eqnarray}
v&=&f(T),\qquad (f(0)=0),\\
X&=&\int_0^T v(T)dT+C_1,\label{3.8} \end{eqnarray} where $C_1$ is
a constant determined by the initial coordinate of the particle.
Then we get the coordinate difference between $A$ and $B$ at time
$T$ from (\ref{3.8}) as
\begin{eqnarray}
 X_A(T)-X_B(T)=X_A(0)-X_B(0)\equiv L . \label{3.9} \end{eqnarray} So the
coordinate difference is a constant in $\S$.

When the speed becomes constant, we can define a comoving
coordinate system $O(t,x,y,z)$ for spaceships $A$ and $B$. Then we
can measure the proper distance between them, namely the rigid
length $L_0 =\overline{AB}$ with local simultaneity $dt=0$. By the
first equation of Lorentz transformation (\ref{2.3}) and
$dt=t_A-t_B=0$, we get
\begin{eqnarray}
T_A-T_B&=&[X_A(T_A)-X_B(T_B)]\tanh\xi =[X_A(T_A)-X_B(T_B)]v \nn\\
&=&[v(T_A-T_B)+X_A(T_B)-X_B(T_B)]v=[v(T_A-T_B)+L ]v,
\end{eqnarray}
and then
\begin{eqnarray}
T_A-T_B=\frac {vL }{1-v^2},\qquad X_A(T_A)-X_B(T_B)=\frac {L
}{1-v^2}.\label{3.11}
\end{eqnarray}
Again by the second equation of Lorentz transformation
(\ref{2.3}), we get the proper distance,
\begin{eqnarray}
L_0  &=&(x_A-x_B)|_{t_A=t_B}\nn\\
&=& [X_A(T_A)-X_B(T_B)]\cosh\xi - (T_A-T_B) \sinh\xi=\frac {L
}{\sqrt{1-v^2}},\label{3.12}
\end{eqnarray}
(\ref{3.12}) means the string stretches.

The above calculation can be regarded as a standard derivation for
the Lorentz-Fitzgerald contraction of a moving rod. More formally,
we have

{\bf Theorem 3.} {\em For a rod $AB$ has proper length $L_0
=\overline{AB}$, assume the rod is parallel to the $X$-axis of the
coordinate system $\S$, and moves along $X$ at constant speed $v$,
then in $\S$, the  {\bf simultaneous distance} (actually the
coordinate difference) $L $ between $A$ and $B$ is given by
\begin{eqnarray}
{L }=L_0  {\sqrt{1-v^2}},
\end{eqnarray}
which is less than the proper length by a factor $\sqrt{1-v^2}$.}

In the case of the constant acceleration $a$, by (\ref{3.6}) we
have
\begin{eqnarray}
\frac 1 2\ln\frac {1+v}{1-v}+\frac {v}{1-v^2}=2aT.
\end{eqnarray}
The solution $X(T)$ is not integrable, so the trajectory with a
constant acceleration may be not $X=\sqrt{T^2+a^{-2}}+C_0$ as
suggested in some textbooks. This is caused by the invalid
definition of constant acceleration.

Now we discuss a variation of the Bell's paradox. That is, what
speed relation between two spaceships can keep the proper length
$L_0 =L $. Similar to the derivation of (\ref{3.11}), substituting
\begin{eqnarray}
X_A(T)=\int_0^T v_A dT+L ,\qquad X_B(T)=\int_0^T v_B
dT\end{eqnarray} into the first equation of Lorentz transformation
(\ref{2.3}), we have the coordinate relations under $t_A=t_B$
\begin{eqnarray}
T_A-T_B = \frac{[X_A(T_B)-X_B(T_B)]v}{1-v^2}=\frac {v}
{1-v^2}\l(\int_0^{T_B}(v_A-v_B) dT +L  \r),\\
X_A(T_A)-X_B(T_B)=\frac {1} {1-v^2}\l(\int_0^{T_B}(v_A-v_B) dT +L
\r),
\end{eqnarray}
Again substituting the above relations into  the second equation
of (\ref{2.3}), we get the proper length $L_0 $ when two
spaceships reach the same speed $v_A=v_B=v=\tanh\xi$ as
\begin{eqnarray}
L_0  &=& [X_A(T_A)-X_B(T_B)]\cosh\xi - (T_A-T_B) \sinh\xi\\
&=&\frac {1} {\sqrt{1-v^2}}\l(\int_0^{T_B}(v_A-v_B) dT +L  \r).
\end{eqnarray}
By $L =L_0 $, we get the constraint on the speed of the two
spaceships
\begin{eqnarray}
\int_0^{T_B}(v_B-v_A) dT =(1- \sqrt{1-v^2})L_0 .
\end{eqnarray}

\section{the paradoxes for rigid body}
\setcounter{equation}{0}
\subsection{the Ladder Paradox}

It is well known, we can not define a rigid body in the framework
of relativity. The consistent matter model should be 4-dimensional
fields\cite{matt1,matt2}. However rigid model can be discussed in
an alternative method similar to the discussion of the Bell's
spaceships. That is, we analyze the motion of the front and back
points of a rod with some suitable elasticity.

The principle of relativity implies a useful manipulation which is
easily overlooked, that is, {\bf the solution to a physical
problem in one coordinate system can be transformed into the
solution in other coordinate system via simultaneous
transformation of coordinates and variables}. To analyze the
coordinate relation between the moving ladder and the garage, we
only need to calculate the trajectories of the front and back
points in the coordinate system with global simultaneity, then we
can get the corresponding results in other coordinate system by
Lorentz transformation, and the results should be the same
obtained by practical measurement in the moving reference frame.

Assume in the coordinate system $S(T,X,Y,Z)$ the garage is static
and we have the global simultaneity $dT=0$. The ladder moves in
$X$ direction at speed $u>0$. We label the front and back points
of the ladder respectively by $A$ and $B$. Then the above results
for the linked Bell's spaceships can be employed. By (\ref{3.8}),
we have the coordinates of $A$ and $B$ in $\S$
\begin{eqnarray}
X_A(T)=u T +L ,\qquad X_B(T)=u T. \label{4.1}\end{eqnarray} The
proper length of the ladder is $L_0 =L  (1-u^2)^{-\frac 1 2}$.
Denote the front and back door of the garage by $C$ and $D$, then
we have coordinates of the door
\begin{eqnarray}
X_C(T)=L _1,\qquad X_D(T)=0.\label{4.2} \end{eqnarray} In the
point of view of the observer in $\S$, if $L _1>L $, we can close
the door $C$ at $T\le \frac 1 u (L _1-L ) $, and close the door
$D$ at $T\ge 0 $. So the garage contains the ladder within the
time
\begin{eqnarray}
0\le T\le \frac 1 u (L _1-L ), \label{4.2.0}
\end{eqnarray} no matter whether $L _1<L_0 $ or not.

Now we examine the two events of closing doors $C$ and $D$ in the
comoving reference frame $\O$ moving at speed $V=u$. Substituting
(\ref{4.1}) and (\ref{4.2}) into the Lorentz transformation
(\ref{2.3}), we get coordinates in $\O$ as follows
\begin{eqnarray}
x_A(T) = \frac {L }{\sqrt{1-u^2}},&&\qquad x_C(T) = \frac {L _1-uT}{\sqrt{1-u^2}}, \label{4.3}\\
t_A(T) = \sqrt{1-u^2}T-\frac{uL }{\sqrt{1-u^2}},  && \qquad t_C(T)
= \frac {T-uL _1}{\sqrt{1-u^2}}, \label{4.4}\end{eqnarray} and
\begin{eqnarray}
x_B(T) = 0,&&\qquad x_D(T) = \frac {-uT}{\sqrt{1-u^2}}, \label{4.5}\\
t_B(T) = \sqrt{1-u^2}T, && \qquad t_D(T) = \frac
{T}{\sqrt{1-u^2}}. \label{4.6}
\end{eqnarray}
By (\ref{4.3}), we find if $ T \le \frac 1 u (L _1-L ) $, we have
$x_A\le x_C$, so we can also close the door $C$. Again by
(\ref{4.5}), we find if $ T\ge 0$, we have $x_B\ge x_D$, then we
can close the door $D$. The combination of the conditions is just
(\ref{4.2.0}). However, $t_C(T)\ne t_D(T)$, which means the doors
are closed at different moment in $\O$. If we want to close the
doors synchronously in $\O$, namely $t_C(T_C)= t_D(T_D)$, then we
get $T_C=T_D+uL _1$. Substituting it into (\ref{4.2.0}) we find if
\begin{eqnarray}
L _1 \ge  \max(L,\frac {uL }{{1-u^2}}),\qquad 0 \le T_D<T_D+uL
_1=T_C\le \frac 1 u (L _1-L ), \label{4.7}
\end{eqnarray}
the garage contains the ladder simultaneously in $\O$.

The above example shows how the kinematics in the Minkowski
space-time is logically consistent with different coordinate
system. {\bf The Minkowski space-time is an evolving
manifold}\cite{gu,Ellis2}, for which only in one special inertial
reference frame we have the global simultaneity. For any physical
process, we can solve the dynamical and kinematical equations in
one convenient coordinate system, and then transform the solution
into the one in the other coordinate system by the principle of
relativity. Since the transformation between coordinates and
variables are 1-1 linear mapping, it certainly does not lead to
contradiction.

The Supplee's submarine paradox is also a paradox for rigid
body\cite{supp}. But it is a dynamical problem involving the
buoyant force and hydrodynamics, so the detailed calculation will
be much complicated. However, we can also remove the paradox by
the principle of relativity. If we get a solution in one
coordinate system, we can transform the solution into the one
described in other coordinate system via linear coordinate and
variable transformations. The transformed solution is identical to
the one solved in the new coordinate system.

\subsection{the Ehrenfest Paradox}

The Ehrenfest paradox may be the most basic phenomenon in
relativity that has a long history marked by controversy and which
still gets different interpretations published in peer-reviewed
journals\cite{rotat}. The paradox involves the curvilinear
coordinate system and global simultaneity. Most of the following
calculation and analysis were given in \cite{gu}, but we introduce
them here for the integrality and clearness of the illustration.

The coordinate transformation between the rotating cylindrical
coordinate system (the Born chart) $B(t,r,\phi,z)$ and the static
Cartesian chart $S(T,X,Y,Z)$ is given by\cite{disc,Born1,Born2}
\begin{eqnarray}
T=t,\quad X=r\cos(\phi+\om t),\quad Y= r\sin(\phi+\om t),\quad
Z=z.  \end{eqnarray} The line element (\ref{2.1}) in
$B(t,r,\phi,z)$ becomes
\begin{eqnarray}
ds^{2}=( 1-{r}^{2}{\om}^{2} )  dt^{2}-2{r}^{2}\om dt d\phi
-dr^{2}-{r}^{2} d\phi^{2}-dz^{2}. \label{4.15}\end{eqnarray} It
includes the `cross-terms' $dt d\phi$, so the time-like vector
$\pa_t$ is not orthogonal to the spatial one $\pa_\phi$. This is
the key of the paradox. In the perspective of the observer in
static coordinate system $\S$, the simultaneity means $dT=0$,
which leads to $dt=0$, and then the spatial length element is
given by
\begin{eqnarray}
dl=|ds|_{dT=0}=\sqrt{dr^2+r^2d\phi^2+dz^2}.
\label{4.16}\end{eqnarray} $dl$ is identical to the length in the
static cylindrical coordinate system, so the circumference at
radius $R$ is still $2\pi R$ rather than $2\pi
R\sqrt{1-R^2\om^2}$. We can not analyze the kinematics in the
curvilinear coordinate system in the point of view for inertial
Cartesian coordinate system. We must use more general concepts and
relations.

However, for an idealized clock attached to the rotational disc,
the proper time is different from that at rest. In this case
$r|\om|<1$, by $dr=d\phi=dz=0$, we get the proper time element of
the moving clock as
\begin{eqnarray}
d\tau = |ds|_{dl=0}=\sqrt{1-{r}^{2}{\om}^{2}}dt.
\label{4.17}\end{eqnarray} (\ref{4.17}) shows the relativistic
effect, the moving clock slows down. In the region $r|\om|\ge 1$,
the Born coordinate is still a 1-1 smooth transformation in the
region $|\phi|<\pi$, so it is still valid. However we can not fix
a clock in the rotational disc $B(t,r,\phi,z)$, because in this
case $ds$ is imaginary, which means $dt$ becomes space-like
element. This is one difference between mathematical coordinate
and physical process.

In the  perspective of an observer attached at $(r_0,\phi_0,z_0)$
in the rotational coordinate system, the spatial vector bases
$(\pa_r,\pa_\phi,\pa_z)$ are also orthogonal to each other, and
then the line element between two local events should take the
form $\dl s^2=\dl t^2-\dl r^2- g_{\phi\phi}\dl\phi^2-\dl z^2$,
where $\dl t$ is his local time element. By the universal
expression of line element (\ref{4.15}), we have
\begin{eqnarray}
ds^{2}=\dl t^{2}-dr^{2}-\frac {r^2}{1-r^2\om^2} d\phi^{2}-dz^{2},
\label{2.7*}\end{eqnarray} where $r=r_0$ for this specified
observer and $\dl t$ is given by
\begin{eqnarray}
\dl t = \sqrt{1-r^2\om^2}dt-\frac
{r^2\om}{\sqrt{1-r^2\om^2}}d\phi. \label{2.8*}\end{eqnarray} The
simultaneity of this observer means $\dl t|_{r=r_0}=0$. For all
observer attached in the rotational coordinate system, the induced
Riemannian line element in the quotient spatial manifold $(r,\phi,
z)$ is generally given by
\begin{eqnarray}
dl^{2}=dr^{2}+\frac {r^2}{1-r^2\om^2} d\phi^{2}+dz^{2},\quad
(r|\om|<1), \label{2.9*}\end{eqnarray} which corresponds to the so
called Langevin-Landau-Lifschitz metric.

Since the 1-form (\ref{2.8*}) is not integrable, so we can not
define a global time for all observers attached in the rotational
coordinate system. The physical reason is that the underlying
manifold of the Born chart is still the original Minkowski
space-time. If we toughly redefine a homogenous time $\wt t$
orthogonal to the hyersurface (\ref{2.9*}), then we get a new
curved space-time equipped with metric
\begin{eqnarray}
ds^{2}=d {\wt t}^{~2}-dr^{2}-\frac {r^2}{1-r^2\om^2}
d\phi^{2}-dz^{2},~~(r|\om|<1). \label{2.10*}\end{eqnarray}
Needless to say, (\ref{2.10*}) and (\ref{2.7*}) describe different
space-time, because they are not globally equivalent to each
other. However, they are equivalent locally.

Generally speaking, the Lorentz transformation is only a local
manipulation in the tangent space-time of a manifold, and the
global Lorentz transformation only holds between Cartesian charts
in the Minkowski space-time\cite{gu1}. In the curved space-time,
by introducing the vierbein and separating the temporal
component\cite{gu4}, the vector and transformation can also be
locally expressed in the form of Clifford algebra (\ref{*2.2}) or
(\ref{*2.3}) with varying $\Om$.

\section{some remarks}
\setcounter{equation}{0}

The above analysis shows how a subtle concept could be
misinterpreted. Such situation also exists in quantum theory. The
Einstein's relativity is essentially a geometrical theory of the
space-time, so it will be much understandable and effective to
directly treat it as Minkowski geometry similar to the Euclidean
one. The postulate of constant light velocity involves a number of
physical concepts and processes without explicit definitions and
explanations, which leads to unnecessary confusions for the one
who is unfamiliar with the details of electrodynamics and wave
equations.

Except for the requirement that the physical equations have the
same form in all admissible frames of reference, some of the
profound implications of the principle of relativity are usually
overlooked in textbooks. One is that, if we get the solution to a
physical process in one coordinate system, then we actually get
the solution in all coordinate system, because the solution can be
uniquely transformed from one coordinate system to the other
though two fold transformation of coordinates and variables. The
coordinates are just artificial labels. The coordinate
transformation can never lead to contradictions if the
transformation is invertible and smooth enough.

The compatible matter model in the Minkowski space-time should be
fields. The classical concepts such as particles and rigid body
are only mathematical idealization for convenience of treatments.
In fundamental level, it is much natural to take a particle as a
special state of a field\cite{gu1,matt1,matt2}. It is also more
reliable and natural to treat the space-time and fields as two
different physical systems with mutual interaction.

There are two special coordinate systems for a concrete physical
system. They are the Gaussian normal coordinate system with global
simultaneity, and the comoving ones with respect to the particles.
All other coordinate systems are mainly helpful and convenient in
mathematics. From this perspective, some concepts such as the
Lorentz transformation of temperature may be meaningless in
physics, because we can only well define temperature in the
comoving coordinate system.

\section*{Acknowledgments}
The paper is a development of arxiv:0901.0309 according to the
advise of Prof. G. Ellis.  The author is grateful to his
supervisor Prof. Ta-Tsien Li for his encouragement and Prof. Max
Tegmark for his kind suggestions.

\end{document}